\magnification1150

\rightline{KCL-MTH-01-30}
\rightline{hep-th//y0107181}

\vskip .5cm
\centerline
{\bf Kac-Moody Symmetries of IIB Supergravity }
\vskip 1cm
\centerline{ Igor Schnakenburg and Peter West }
\vskip .5cm
\centerline{Department of Mathematics}
\centerline{King's College, London, UK}

\leftline{\sl Abstract}
\noindent 
We formulate  the bosonic sector of  IIB supergravity as a
non-linear realisation. We show that this non-linear realisation contains
the Borel subalgebras of SL(11) and $E_7$ and  argue that it can be
enlarged so as to be   based on the rank eleven Kac-Moody algebra
$E_{11}$ 
\vskip .5cm

\vfill

\vskip 1cm
email: schnake, pwest@mth.kcl.ac.uk

\eject

%\pageno=1

\medskip
{\bf {0. Introduction }}
\medskip
It is well known since the work of Nahm [1] that 
there is only one supergravity  theory in eleven dimensions, but
there are two maximally supersymmetric theories in ten dimensions.  By
compactifying  one of the  dimensions of the eleven dimensional 
supergravity theory [2], one
finds one of the maximal supergravity theories in ten dimensions. 
This theory has a supersymmetry whose 32 component supercharge is a
Majorana spinor and it has become known as  the  IIA supergravity theory
[3].  The other maximal supergravity theory in ten dimensions, called IIB
supergravity [4,5,6], has two  supercharges that are two 16 component
Majorana-Weyl spinors of same chirality. 
 By virtue of their high degree of supersymmetry and corresponding
uniqueness, the maximal  supergravity theories in ten dimensions  are 
the low energy effective actions for the corresponding
superstring theories.  As such,  they can provide important information
about these string theories including some aspects of their
non-perturbative behaviour.
\par
One of the most remarkable features of supergravity theories is that 
the scalars they contain always occur in a coset structure. While this 
can be viewed as a consequence of supersymmetry, the groups that occur in
these cosets are rather mysterious. The two most studied examples are
perhaps the $E_7$/$SU(8)$ [7] of the maximal supergravity in four
dimensions and  the $SU(1,1)$/$U(1)$ [4] of the ten dimensional
IIB theory. 
It has been conjectured [8] that the symmetries found in these  cosets are
symmetries of the associated  non-perturbative string theory. 
\par
The coset construction was extended [9] to include the gauge fields of
supergravity theories. This method used generators that were inert
under Lorentz transformations and, as such, it is difficult to extend
this method to include either gravity or the fermions. However, this
construction did include the  gauge and scalar fields as well as
their duals, and as a consequence  the  equations of motion for these
fields could be expressed as a generalised self-duality condition.
This formulation was given for eleven dimensional supergravity, all
its reductions to four dimensions as well as for the IIB theory
[9]. 
\par
Recently [10], it was shown that the entire bosonic sectors  of eleven
dimensional and ten dimensional IIA supergravity theories could be
formulated as  non-linear realisations. 
In this way of proceeding gravity and the gauge fields appeared on an
equal footing and one could hope to see the full symmetries of
supergravity theories. 
\par
Here we shall confirm the  conjecture in references [10,11] that the
entire bosonic sector of ten dimensional IIB supergravity can also be
formulated as a non-linear realisation.  The formulation of the IIB
theory we find is one in which  all the degree of freedom  of the
theory,  except for the graviton and the four form gauge field which
satisfies a self-duality condition, are described by a gauge field and
its dual gauge field. The equations of motion just relate the two
field strengths using Hodge duality.  This construction is carried out in
section one. 
\par
In reference [11] it was argued that the non-linear realisation of eleven
dimensional supergravity given in [10] could be extended so   that
eleven dimensional supergravity could be formulated as a non-linear
realisation based on a rank eleven  Kac-Moody algebra called $E_{11}$
in [11]. In section two we assume that a similar
enlargement is possible for the non-linear  realisation of the IIB theory
given in  section one. By first showing  that the non-linear
realisation of   section one  contains 
the Borel subalgebras  of SL(11) and $E_7$,  we identify this
Kac-Moody algebra to be also $E_{11}$. We then outline how this non-linear
realisation can be enlarged to contain the Borel subalgebra of $E_8$.
This last step implies that the gravitational degrees of freedom must be
described by by two fields which are related by duality. Finally, we
discuss the consequences of these results for the relationship between
the IIA and IIB theories and M theory. 
\medskip
{\bf 1. IIB Supergravity as a Non-linear Realisation}
\medskip
The degrees of freedom of this theory consists of  the graviton, denoted
by the field ${h_a}^b$, two scalars  $A^1$ and $A^2$ corresponding to
the dilaton  and the axion respectively,  two 2-form gauge potentials 
$A^1_{a_1 a_2}$ and $A^2_{a_1 a_2}$ and a 4-form gauge potential 
$A_{a_1 \ldots a_4}$. The graviton belongs 
to the NS-NS sector in the associated IIB string theory as do the other
fields if they carry the superscript  one. The fields carrying the
superscript two  belong to the R-R sector.   
\par
As in references  [9] and [10], we introduce  dual fields for all
the fields except for the graviton and the four form gauge field, whose
field strength satisfies a type of  self-dual condition. The
complete  field content is then given by the set 
$$
   h_a{}^b, A^s, \,  A_{c_1 c_2}^s, \, A^2_{c_1 \ldots c_4}, ,\ A_{c_1
     \ldots  c_6}^s,\, A_{c_1 \ldots c_8}^s,
\eqno(1.1)
$$
where $s$ can take value 1 or 2 corresponding to the sectors. Each of
the above fields is to be a Goldstone boson and as such we introduce
a corresponding set of generators which is then given by  
$$
 K^a{}_b,\, R_s, \,  R^{c_1 c_2}_s,\, R_2^{c_1 \ldots c_4},\, R^{c_1
       \ldots c_6}_s,\,  R^{c_1 \ldots c_8}_s,\qquad s = 1,\, 2
\eqno(1.2)
$$
We also include the momentum generator $P_c$ which  introduces space-time
into the group element. 
\par
We take the  generators to obey the following relations:
$$
   [{K^a}_b ,  {K^c}_d] = \delta^c_b{K^a}_d -\delta^a_d{K^c}_b  ,
   \quad [{K^a}_b , P_c] = - \delta^a_cP_b, 
$$
$$
   [{K^a}_b  ,   R_s^{c_1\cdots c_p}] = \delta^{c_1}_bR_s^{ac_2\cdots
   c_p} + \cdots  , 
$$
$$
   [R^{c_1\cdots c_p}_{s_1}    ,   R^{c_1\cdots c_q}_{s_2}] = c_{p,q}^{s_1,
   s_2}R^{c_1\cdots c_{p+q}}_{s(s_1,s_2)}, 
$$
$$
   [R_1, R_s^{c_1\cdots c_p} ] = d_p^s R^{c_1\cdots c_p}_s,\quad [R_2,
   R_{s_1}^{c_1\cdots c_p} ] = \tilde d_p^s R^{c_1\cdots c_p}_{s(2,s_1)},
\eqno(1.3)
$$
where $ + \cdots $ means the appropriate anti-symmetrisations.
The generators ${K^a}_b$ satisfy the
commutation relations of $GL(10,{\bf{R}})$. In the third line the
superscript $s$ depends on the  fields in the commutator and we have
therefore written $s = s(s_1,s_2)$. This function satisfies
the properties $s(1,1) = s(2,2) = 1, \quad s(1,2) = s(2,1) = 2$. 
In the last line we have split the scalar commutators into those
for the dilaton (superscript 1) with coefficient $d_p^s$, and the
axion (subscript 2) with coefficient $\tilde d_p^s$. One can see
that in the commutator the dilaton is sector preserving while the axion
changes the sector of the other generator. The Jacobi identity implies the
following relations among the constants. 
$$
   c_{q,r}^{s_2,s_3}c_{p, q+r}^{s_1, s(s_2,s_3)} =
   c_{p,q}^{s_1,s_2}c_{p+q, r}^{s(s_1,s_2),s_3} +
   c_{p,r}^{s_1,s_3}c_{q, p+r}^{s_2,s(s_1,s_3)} ,
\eqno(1.4)
$$
where 
$$
c_{0,q}^{2,s_2} \equiv \tilde d^{s_2}_p
$$
and 
$$
(d_{q}^{s_3} + d_{p}^{s_1} - d_{p+q}^{s(s_1,s_3)}) c_{p,q}^{s_1,s_3}=0.   
\eqno(1.5)
$$
The constants in the above commutation relations are taken to be:
$$
   d_2^1 = d_6^2 = -d_2^2 = -d_6^1 = {1 \over 2},\   d_0^2 =
   -d_8^2 = -1,
\eqno(1.6)
$$
$$ 
   c_{2,2}^{1,2} = - c_{2,2}^{2,1} = -1 , \quad  
   c_{2,4}^{2,2}  = - c_{2,4}^{1,2}  = 4 , \quad
   c_{2,6}^{1,2}  = 1  ,\quad
   c_{2,6}^{1,1}  = - c_{2,6}^{2,2} = {1\over 2}
$$
$$ 
   \tilde d_2^1 = - \tilde d_{6}^{2} = - \tilde d_{8}^{2}  = 1, \quad
   \tilde d_2^2 = \tilde d_{6}^{1}  = \tilde d_8^1 = 0 
\eqno(1.7)
$$
All not mentioned coefficients are zero. One can verify that they do
indeed satisfy the Jacobi relations. We denote the above algebra by
$G_{IIB}$. 
\par
The algebra $G_{IIB}$ possesses the Lorentz algebra as a subalgebra. 
The generators $J_{ab}$ of the latter are given by the anti-symmetric
part of the ${K^a}_b$ generators, i e: $J_{ab} = K_{ab}- K_{ba}$,
where the indices are lowered and raised with the Minkowski metric. 
We will show that IIB supergravity can be described as a nonlinear
representation of the group $G_{IIB}$ taking the Lorentz group as the
local subgroup. The general element of $G_{IIB}$ can be written as 
$$
   g = \exp (x^\mu P_\mu) \exp ({h_a}^b{K^a}_b) g_A\equiv g_h g_A, 
   \eqno(1.8)
$$
where
$$
   g_A = e^{(1/8!)A^2_{a_1\cdots a_8} R^{a_1\cdots a_8}_2} 
   e^{(1/8!)A^1_{a_1\cdots a_8} R^{a_1\cdots a_8}_1} 
   e^{(1/6!)(A^2_{a_1\cdots a_6}R^{a_1\cdots a_6}_2
   + A^1_{a_1\cdots a_6}R^{a_1\cdots a_6}_1)}
$$
$$
   \times  e^{(1/4!)A^2_{a_1\cdots a_4}R^{a_1\cdots a_4}_2} \,
   e^{(1/2!)(A^2_{a_1a_2}R^{a_1a_2}_{2} +
   A^1_{a_1a_2}R^{a_1a_2}_{1})} \, e^{A^2 R_2} \, e^{A^1 R_1}.
\eqno(1.9)
$$ 
For easier identification with the known literature in what follows below 
we will often relabel $A^1=\sigma$ and $A^2=\chi$. 
\par
Following the standard procedure of non-linear realisations we demand
that the theory be invariant under 
$$
   g \to g_0 g h^{-1},
\eqno(1.10)
$$
where $g_0$ is an element from the whole group $G_{IIB}$ and is a rigid
transformation while $h$ is a local Lorentz transformation.
\par
We now calculate the Maurer-Cartan form 
$$
   {\cal {V}} = g^{-1}dg -\omega
\eqno(1.11)
$$
in the presence of the Lorentz connection $\omega ={1\over 2}
dx^\mu\omega_{\mu b}{}^a J^b{}_a$, which transforms as
$$
   \omega \to h\omega h^{-1} + hdh^{-1}.
\eqno(1.12)
$$
As a result, ${\cal {V}}$ transforms as 
$$
   {\cal {V}} \to h {\cal {V}}  h^{-1}.
\eqno(1.13)
$$
Writing ${\cal {V}}$ in the form 
$$
   {\cal {V}} = (g_h^{-1}dg_h ) + (g_A^{-1}dg_A 
+ g_A^{-1}(g_h^{-1}dg_h)g_A -
   g_h^{-1}dg_h),
\eqno(1.14)
$$
using the relations 
$$
   e^{-A} d e^A  = dA - {1\over 2} [A, dA] + {1\over 6} [A, [A, dA]] -
   {1\over 24} [A, [A , [A ,dA]]] + \cdots ,
$$
$$
   e^{-A} B e^A  = B - [A, B] +{1\over 2} [A, [A, B]] + \cdots ,
\eqno(1.15)
$$ 
and  the commutation relations  of the $G_{IIB}$ algebra  and we find
that  
$$
   {\cal {V}} \equiv dx^\mu (e_\mu{}^a P_a + dx^\mu {\Omega_{\mu
   a}}^b{K^a}_b) + dx^\mu ( \sum_{p=1}^8{1\over
p!}e^{-d_{p-1}^{s}\sigma}\tilde D_\mu
   A_{a_1\cdots a_p}^sR^{a_1\cdots a_p}_s),
\eqno(1.16)
$$
where 
$$
   e_\mu{}^a= (e^h)_\mu{}^a,
\eqno(1.17)
$$
$$
   {\Omega_{ab}}^c \equiv {(e^{-1})_a}^\mu {(e^{-1}\partial_\mu e)_b}^c -
   {\omega_{ab}}^c,
\eqno(1.18)
$$
and the definition of  $\tilde D_\mu A_{a_1\cdots a_p}$ will be given
below. 
\par
The IIB supergravity theory is the non-linear realisation of 
the group that is the closure of the $G_{IIB}$ algebra given above with
the ten dimensional conformal algebra. However, rather than working with
this infinite dimensional group we first construct the Cartan forms of the
$G_{IIB}$ algebra, as above, and then take only such combinations of
these that can be rewritten in terms of Cartan forms of the
conformal group. This procedure was described in detail in
reference [10] and we will simply state the results of this method
when applied to the $G_{IIB}$ algebra. In the gravity sector we adopt the
unique constraint 
$$
   \Omega_{a[bc]} - \Omega_{b(ac)} + \Omega_{c(ab)}=0,
\eqno(1.19)
$$
which gives the usual expression for the spin connection in terms of
the vielbein. The only objects which are  Lorentz covariant and therefore
covariant under the full non-linear realisation composed out of the
closure of the conformal and $G_{IIB}$ algebras are the Riemann tensor
composed out of the spin-connection in the usual way and the completely
anti-symmetrised derivatives $e^{-d_{p-1}^{s}\sigma}\tilde
D_{[a_1}A_{a_2\cdots a_p]} $. The latter are denoted by 
$$
   \tilde F_{a_1\cdots a_p}^s = pe^{-d_{p-1}^{s}\sigma}\tilde
   D_{[a_1}A_{a_2\cdots a_p]}. 
\eqno(1.20)
$$
We observe that these expressions begin with the field strength of the
gauge fields as they should. The
explicit expressions for these objects, whose calculation was explained
above, are then  given for the scalars by 
$$
 \tilde   F_a^1 = \tilde D_a \sigma ,\qquad
 \tilde   F_a^2 = e^{\sigma}\tilde D_a \chi,
\eqno(1.21)
$$
for the 2-index fields:
$$
 \tilde   F^1_{a_1a_2a_3}  = 3e^{-{1\over 2}\sigma} \tilde
D_{[a_1}A^1_{a_2a_3]},\quad
 \tilde   F^2_{a_1a_2a_3} = 3e^{{1\over 2}\sigma} 
   ( \tilde D_{[a_1}A^2_{a_2a_3]} - \chi
   \tilde D_{[a_1}A^1_{a_2a_3]}), 
\eqno(1.22)
$$
for the 4-index field 
$$
 \tilde   F_{a_1\cdots a_5}  = 5 ( \tilde D_{[a_1}A_{a_2\cdots a_5]} + 3
   A^1_{[a_1a_2}\tilde D_{a_3}A^2_{a_4a_5]}
   -3A^2_{[a_1a_2}\tilde D_{a_3}A^1_{a_4a_5]} ),
\eqno(1.23)
$$
for  the 6-index fields
$$
   \tilde F^2_{a_1\cdots a_7}  = 7e^{-{1\over 2}\sigma}\left(
   \tilde D_{[a_1} A^2_{a_2\cdots
   a_7]} + 60 A^1_{[a_1a_2} ( \tilde D_{a_3}A^2_{a_4\cdots a_7]} +
   A^1_{a_3a_4}\tilde D_{a_5}A^2_{a_6a_7]} -
   A^2_{a_3a_4}\tilde D_{a_5}A^1_{a_6a_7]})\right),
\eqno(1.24)
$$
$$ 
   \tilde F^1_{a_1\cdots a_7}  = 7e^{{1\over 2}\sigma}\left(
   \tilde D_{[a_1} A^1_{a_2\cdots
   a_7]} - 60 A^2_{[a_1a_2} ( \tilde D_{a_3}A^2_{a_4\cdots a_7]} +
   A^1_{a_3a_4}\tilde D_{a_5}A^2_{a_6a_7]} -
   A^2_{a_3a_4}\tilde D_{a_5}A^1_{a_6a_7]}) \right)
$$
$$ 
   + 7 e^{{1\over 2}\sigma}\chi \left( 
   \tilde D_{[a_1} A^2_{a_2\cdots
   a_7]} + 60 A^1_{[a_1a_2} ( \tilde D_{a_3}A^2_{a_4\cdots a_7]} +
   A^1_{a_3a_4}\tilde D_{a_5}A^2_{a_6a_7]} -
   A^2_{a_3a_4}\tilde D_{a_5}A^1_{a_6a_7]})\right),
\eqno(1.25) 
$$
and finally for the 8-index fields
$$
   \tilde F^1_{a_1\cdots a_9} = 9 \left(  \right.
   \tilde D_{[a_1} A^1_{a_2\cdots
   a_9]}- 7\cdot 2 A^1_{[a_1a_2} \left(  \right.
   \tilde D_{a_3}A^1_{a_4\cdots a_9]} - 6\cdot 5 A^2_{a_3a_4}(
   \tilde D_{a_5}A^2_{a_6\cdots a_9]} 
$$
$$
   + {1\over 2}A^1_{a_5a_6}\tilde D_{a_7}A^2_{a_8a_9]} -
   {1\over 2} A^2_{a_5a_6}\tilde D_{a_7}A^1_{a_8a_9]}
   )\left. \right) + 7\cdot
   2  A^2_{[a_1a_2} \left( \right.
   \tilde D_{a_3}A^2_{a_4\cdots a_9]} + 6\cdot 5 A^1_{a_3a_4}(
   \tilde D_{a_5}A^2_{a_6\cdots a_9]}
$$
$$  
   + {1\over 2}A^1_{a_5a_6}\tilde D_{a_7}A^2_{a_8a_9]} - {1\over 2}
   A^2_{a_5a_6}\tilde D_{a_7}A^1_{a_8a_9]})
   \left.\right)
   \left. \right) + 9\chi\left(
    \right.
   \tilde D_{[a_1}A^2_{a_2\cdots a_9]}
$$
$$
   - 7\cdot 4 A^1_{[a_1a_2}
   \left( \right.
   \tilde D_{a_3}A^2_{a_4\cdots a_9]} + 6\cdot 5 A^1_{a_3a_4}(
   \tilde D_{a_5}A^2_{a_6\cdots a_9]} 
   +{1\over 2}A^1_{a_5a_6}\tilde D_{a_7}A^2_{a_8a_9]} -
   {1\over 2} A^2_{a_5a_6}\tilde D_{a_7}A^1_{a_8a_9]}
   )\left.\right) \left.\right)
\eqno(1.26)
$$
and
$$
   \tilde F^2_{a_1\cdots a_9} = 9 e^{-\sigma}\left(\right.
   \tilde D_{[a_1}A^2_{a_2\cdots a_9]}- 
   7\cdot 4 A^1_{[a_1a_2}
   \left( \right.\tilde D_{a_3}A^2_{a_4\cdots a_9]} + 6\cdot 5 A^1_{a_3a_4}(
   \tilde D_{a_5}A^2_{a_6\cdots a_9]}
$$
$$ 
   +{1\over 2}A^1_{a_5a_6}\tilde D_{a_7}A^2_{a_8a_9]} - {1\over 2}
   A^2_{a_5a_6}\tilde D_{a_7}A^1_{a_8a_9]} )\left.\right)
\left.\right) .
\eqno(1.27)
$$
\par
The equations of motion can only be constructed from the spin
connection and the covariant objects $\tilde  F^s_{a_1\ldots a_p}$ and
can only be 
$$
  \tilde  F^{1 \mu\nu\rho}  = {1\over 7!}\epsilon^{\mu\nu\rho\mu_1\cdots
   \mu_7} \tilde F^1_{\mu_1\cdots\mu_7}, \quad \tilde  F^{2 \mu\nu\rho}
   = {1\over 7!}\epsilon^{\mu\nu\rho\mu_1\cdots\mu_7} \tilde
   F^2_{\mu_1\cdots\mu_7},
\eqno(1.28)
$$
$$
  \tilde  F^{1 \mu} = {1\over 9!}\epsilon^{\mu\mu_1\cdots\mu_9}
   \tilde F^1_{\mu_1\cdots\mu_9},\quad 
 \tilde   F^{2 \mu} = {1\over
9!}\epsilon^{\mu\mu_1\cdots\mu_9} \tilde F^2_{\mu_1\cdots\mu_9},
\eqno(1.29)
$$
The remaining equation of motion is that for the vielbein and is given
by 
$$
   R_{\mu\nu} - {1\over 2} g_{\mu\nu} R -  [
   {1 \over 2}\partial_{(\mu}\sigma\partial_{\nu)}\sigma + 
   {1\over 2}e^{2\sigma}\partial_{(\mu}\chi\partial_{\nu)}\chi 
   - {1\over 4} g_{\mu\nu}\left( \partial_{\mu}\sigma\partial^{\mu} \sigma + 
   e^{2\sigma}\partial_{\mu}\chi\partial^{\mu}\chi \right) +
$$
$$
   - {1\over 6}\tilde F_{\mu_1\cdots\mu_4\mu}
\tilde F^{\mu_1\cdots\mu_4}{}_\nu -
   {1\over 16} e^\sigma
\tilde F_{(\mu}^{2 \mu_1\mu_2}\tilde F^2_{\nu)\mu_1\mu_2} -
   {1\over 16} e^{-\sigma}
\tilde F_{(\mu}^{1 \mu_1\mu_2}\tilde F^1_{\nu)\mu_1\mu_2} +
$$
$$
   + {1\over 96}
   g_{\mu\nu}(e^\sigma \tilde F^{2 \mu_1\mu_2\mu_3}
\tilde F^2_{\mu_1\mu_2\mu_3}
+
   e^{-\sigma} \tilde F^{1 \mu_1\mu_2\mu_3}\tilde F^1_{\mu_1\mu_2\mu_3})
] = 0
\eqno(1.30)
$$
The value of the constants in front of the field strength squared terms
 can only be fixed by considering the full  non-linear realisation  of 
the IIB theory that includes the fermionic sector of the theory or the
Kac-Moody  groups considered later in this paper.  In the above equation
we have fixed the values of these constants to their correct values. 
\par
We can obtain the more standard second order equations of IIB supergravity
in terms of the original fields without their duals   by differentiating
the equations (1.28) and (1.29) and using the Bianchi identities of the
dual field strengths.  For example, 
 we can rewrite the second equation in equation (1.28) as 
$\tilde F^2_{\mu\nu\rho} = {1\over
7!}\epsilon_{\mu\nu\rho\mu_1\cdots\mu_7}
\tilde F^{2 \mu_1\cdots\mu_7}$, inserting  expressions (1.22) and
(1.24) for these field strengths into this equation, bringing the
exponential of
$\sigma$  to the left hand side, we find that differentiating  the whole
expression yields the equation 
$$
   \partial^\rho ( e^\sigma \tilde F_{\mu\nu\rho}^2 )
   = {2\over 3} \tilde F^{1\rho\mu_1\mu_2}G_{\mu\nu\rho\mu_1\mu_2}.
\eqno(1.31)
$$
Similarly we find that the other equations imply that 
$$
   \partial^\rho ( e^{-\sigma}\tilde  F_{\mu\nu\rho}^1 )  = - {2\over
   3}G_{\mu\nu\rho\lambda\sigma}\tilde  F^{2\rho\lambda\sigma} 
   +\partial^\rho\chi (e^\sigma\tilde  F_{\mu\nu\rho}^2),
$$
$$
   \partial^\mu ( e^{2\sigma}\partial_\mu \chi )  = -{1\over 6} e^\sigma
\tilde   F^{1\mu\nu\rho} \tilde F^2_{\mu\nu\rho} ,
$$
$$ 
   \partial^\mu ( \partial_\mu \sigma ) 
   =e^{2\sigma}\partial^\mu\chi\partial_\mu\chi +{1\over 12}e^\sigma
\tilde   F^{2 \mu\nu\rho}\tilde F^2_{\mu\nu\rho} -{1\over 12}e^{-\sigma}
  \tilde  F^{ 1 \mu\nu\rho}\tilde  F^1_{\mu\nu\rho} .
\eqno(1.32)
$$
Since the  field strength of the four form gauge field
is self-dual we leave this equation to be  first order in derivatives.
\par
The IIB supergravity theory was first discovered [4,5,6] in a
formulation in which the scalars belong to the coset
$SU(1,1)/U(1)$. In this formulation the field content consists of the
graviton, a complex two form gauge fields $A_{a_1a_2}^s$, a real
four form $A_{a_1\ldots a_4}$ and two scalars denoted by the complex
field $\phi$ which belong to the coset $SU(1,1)/U(1)$. The rewriting of
the IIB supergravity theory in terms of an SL(2,{\bf R})/SO(2) coset
was given in reference [12] and was 
reviewed in reference [13]. The equations of motion given above are 
precisely those found in this latter formuation.

%%%%%%%%%%%%%%%%%%%%%%%%%%%%%%%%%%%%%%%%%%%%%%%%%%%%%%%%%%%%%%%%%%
\medskip
{\bf 2. $E_{11}$ and IIB Supergravity}
\medskip
%%%%%%%%%%%%%%%%%%%%%%%%%%%%%%%%%%%%%%%%%%%%%%%%%%%%%%%%%%%%%%%%%%

In reference [11] it was argued that eleven dimensional supergravity 
was invariant under a Kac-Moody algebra that was identified to be 
$E_{11}$. We refer the reader to this paper for a discussion of
how the non-linear realisation of eleven dimensional supergravity 
given in reference [10] might be extended to incorporate such a large
algebra by using an alternative formulation of eleven dimensional
supergravity  and increasing the size of the local subgroup. The same 
group was identified as a symmetry of the IIA supergravity theory. In
this section, we will assume that the non-linear realisation of the IIB
supergravity theory given above can be similarly enlarged to 
a non-linear realisation of a Kac-Moody algebra. We will show   that
this algebra is also $E_{11}$. 
\par
The proposed Kac-Moody algebra of the IIB theory must contain the algebra
denoted 
$G_{IIB}$ above and given in equations (1.3) to (1.7). 
In the non-linear realisation discussed above the local subgroup is taken
to be  the ten dimensional Lorentz group and so all the 
remaining generators in $G_{IIB}$ are coset generators and as such
correspond to fields in  IIB supergravity. In the enlarged non-linear
realisation based on a Kac-Moody algebra, the local subgroup is  taken to
be that invariant under the Cartan involution and as a result  the coset
representatives can be written as exponentials of the Cartan subalgebra
and positive root generators of the Kac-Moody algebra.    Consequently,
all the generators of   $G_{IIB}$, except the negative root generators of
SL(10), must be included in  the Cartan
subalgebra and positive root generators of the Kac-Moody algebra.
A set of commuting generators of $G_{IIB}$ can be taken to be 
$$
   K^a{}_a, \ a=1,\ldots 10, \quad {\rm and}\quad  R_1
\eqno(2.1)
$$
and these may be taken to  belong to  the Cartan subalgebra of the
Kac-Moody algebra.  We also
observe that the remaining generators of $G_{IIB}$, except the negative
root generators of SL(10), can be generated by taking multiple commutators of the generators  
$$ 
   K^a{}_{a+1}, \,\, a=1,\ldots 9,\quad  R_2 \quad {\rm{and }} \quad
   R_1^{910}  .
\eqno(2.2)
$$
We may identify these as positive simple root generators of the
Kac-Moody algebra. Thus we are seeking a rank eleven Kac-Moody algebra. 
\par
Calculating the commutator of the positive simple root and Cartan
sub-algebra generators leads to the Cartan matrix from which we can
uniquely identify the  Kac-Moody algebra. However, working with only
the Cartan sub-algebra and simple positive root generators -that is
without the negative simple  root generators- does not
automatically encode the  particular basis for the Cartan sub-algebra
that satisfies the  Chevalley relations and hence 
produces the correct Cartan matrix. However, we must use a basis that 
leads to an acceptable Cartan matrix, that is one which satisfies
the correct properties to be associated with a Kac-Moody algebra.
Even taking this into account, the choice of the basis is not free
from ambiguity. As in reference [11] this ambiguity may be resolved by
identifying appropriate subgroups and so we  first carry out this step. 
\par
The non-linear realisation of the IIB supergravity theory given above is
obviously invariant under SL(10) generated by $K^a{}_b,\ a,b=1,\ldots 10$,
but as we now show it also is invariant under the Borel subgroup  of
SL(11). One can verify,  using equations (1.3) to (1.7) that the
generators 
$$  
   \hat K^{ a}{}_{ b}, \quad \hat K^a{}_{10}= R_1^{a 10},\quad
   \hat K^a{}_{11}=-R_2^{a10},
$$
$$
   \hat K^{10}{}_{11}=R_2, \quad 
   \hat K^{11}{}_{11}= R_1+{1\over 4}\sum _{a=1}^{10} K^a{}_{a}
   -K^{10}{}_{10}, \quad a,b=1,\ldots 9,
\eqno(2.3)
$$ 
do indeed obey the commutation relations for the Borel subgroup of SL(11). 
We note that this SL(11) only coincides with the SL(9) subgroup of
the obvious SL(10) group, that is for the  generators which carry the
indices $a,b=1,\ldots ,9$.  
\par
The $G_{IIB}$ algebra also contains the rank three anti-symmetric 
representation of the Borel subgroup of SL(11). More precisely, if  we
identify 
$$ 
   X^{a_1a_2a_3}=R_2^{10 a_1a_2a_3},  X^{10 a_1a_2}= R_2^{a_1a_2}, 
   X^{11a_1a_2} = R_1^{ a_1a_2}, X^{a_1 10 11}=K^{a_1}{}_{10},
   a_1,.,a_3  = 1,\ldots , 9
\eqno(2.4)
$$
then equations (1.3) to (1.7) imply the relation 
$$
   [\hat K^{\hat a}{}_{\hat b}, X^{\hat c_1\hat c_2\hat c_3}]=
   3 \delta^{[\hat c_1}_{\hat b} X^{|\hat a|\hat  c_2\hat c_3]}, 
   \ a\le b,\ \hat a, \hat b, \hat c_1, \hat c_2, \hat c_3 = 1 ,\ldots ,
11
\eqno(2.5)
$$ 
\par
To identify the Borel subgroup of $E_7$ contained in $G_{IIB}$ we
consider only the generators whose indices take the
values $i,j,\ldots = 5,\ldots 10$. We search for the formulation
of the $E_7$ algebra which has its SL(7) subgroup manifest. The
generators 
$\hat K^{\hat i}{}_{\hat j},
\ \hat i\le \hat  j$  generate the 27 dimensional Borel sub-algebra of
this SL(7) together with a U(1) factor. The generators 
$X^{\hat i_1\hat i_2\hat i_3}$ belong to the 35 dimensional third rank
anti-symmetric representation of SL(7). Calculating the  commutator of
the later generators we find that 
$$
   [X^{\hat i_1\hat i_2\hat i_3},X^{\hat
i_4\hat i_5\hat i_6}]=2\epsilon^{\hat i_1\ldots \hat i_6 \hat k} 
S_{\hat k}
\eqno(2.6)$$ 
where the 7 generators $ S_k$ are given by 
$$ 
 S_i= {1\over 2.4!} \epsilon_{1011ij_1\ldots j_4}R^{j_1\ldots
j_4}_2,\ 
 S_{10}= {2\over 5 !} \epsilon_{1011j_1\ldots j_5}R_2^{j_1\ldots
j_510},
$$
$$ 
 S_{11}= {2\over 5!} \epsilon_{1011j_1\ldots j_5}R^{j_1\ldots
j_5 10}_1,\ j_1\ldots =5,\ldots, 9. 
\eqno(2.7)
$$
It is then straightforward to verify that the generators 
$$
\hat K^{ \hat i}{}_{\hat j}, \ \hat i\le \hat j,\ X^{\hat
i_1\hat i_2\hat i_3},  S_{\hat k}
\eqno(2.8)
$$ 
satisfy all the remaining commutation relations of the 70 dimensional 
Borel subgroup of
$E_7$. One can also check that for this restriction of the  $G_{IIB}$ 
that there are no other generators contained in $G_{IIB}$ other than the
negative root generators of SL(7). This demonstrates that IIB
supergravity is invariant under the Borel subgroups of SL(11) and
$E_7$. However, we expect that it is also invariant under the full 
SL(11) and
$E_7$ groups, the missing generators forming part of an enlarged local
subgroup. 
\par
Given the above identification of the SL(11) and $E_7$ groups in the
proposed rank eleven Kac-Moody algebra we can finally identify this  
Kac-Moody Lie algebra. The simple positive root generators 
are given by 
$$
   E_a=K^a{}_{a+1}, a=1,\ldots 8,\  E_9=R_1^{9 10},\ E_{10}=R_2,\ 
E_{11}=K^9{}_{10}.
\eqno(2.9)
$$
These agree with the simple positive root generators 
 of the SL(11) and $E_7$ subgroups found above. The
basis of the Cartan subalgebra that leads to an acceptable Cartan matrix
and agrees with the above identifications of the Cartan
subalgebra elements of SL(11) and $E_7$  is given
by 
$$
   H_a=K^a{}_a -K^{a+1}{}_{a+1}, a=1,\ldots, 8,\
H_{9}=K^{9}{}_{9}+K^{10}{}_{10}+R_1-{1\over 4}\sum_{a=1}^{11}K^a{}_a, 
$$
$$
   H_{10}=-2R_1,\ H_{11}=K^{9}{}_{9}-K^{10}{}_{10}
\eqno(2.10)
$$ 
One can
verify that 
$$  
   [H_a,E_b]=A_{ab}E_b
\eqno(2.11) 
$$
where $A_{ab}$ is the Cartan matrix for $E_{11}$. Hence we identify
$E_{11}$ as the Kac-Moody algebra that underlies IIB supergravity. 
\par
In reference [11] it was explained how one might enlarge the
non-linear realisation of eleven dimensional supergravity such
that it contained the $E_{11}$ Kac-Moody algebra. In particular, 
it was shown how by adopting a first order formulation of
gravity involving the usual metric and a dual field one could extend the
non-linear realisation to include the Borel subgroup of $E_8$. 
We refer the reader to this reference for the details of this procedure
and we now outline the analogous steps for IIB supergravity. 
\par
We consider the restriction of the $G_{IIB}$ algebra  resulting from only
considering generators with the indices $i,j=4,\ldots 10$. We will  
find the formulation of the $E_{8}$ algebra with the SL(8) symmetry
manifest and so the generators will carry the  indices $\hat i,\hat
j=4,\ldots 11$. The Borel subgroup of this SL(8) is provided by the
generators 
$\hat K^{\hat i}{}_{\hat j},\ \hat i\le \hat j$. Since this coincides
with only the SL(6) subgroup of the obvious SL(8) we will have to treat
the indices from 
$\hat i,\hat j= 10,11$ differently from $\hat i,\hat j=4,\ldots 9$. The 
$E_8$ algebra possesses the commutation relations 
$$
   [ X^{\hat i_1\hat i_2\hat i_3}, X^{\hat i_4\hat i_5\hat i_6}]
=\epsilon ^{\hat i_1\hat i_2\hat i_3 \hat i_4\hat i_5\hat i_6 \hat k\hat
l} S_{\hat k\hat l}
\eqno(2.12)
$$
and 
$$ 
[ S_{\hat k_1\hat k_2}, X^{\hat i_1\hat i_2\hat i_3}]
=3\delta ^{[\hat i_1\hat i_2}_{\hat k_1\hat k_2} S^{\hat i_3]}
\eqno(2.13)
$$
Calculating the commutators of the generators $X^{\hat i_1\hat i_2\hat
i_3}$ of equation (2.4), we indeed find generators  $S_{\hat k\hat l}$ 
with the exception that $S_{1011}$ is absent. These new generators then 
obey equation (2.13) and lead to new generators $S^{\hat k}$ with the
exception of $S^{i},\ i=4,\ldots 9$. 
\par
The resolution of this dilemma is to modify the $G_{IIB}$ algebra. We
modify the commutation relation 
$$
[R_2^{a_1\ldots a_6}, R_2^{a_7a_8}]={1\over 2}R_1^{a_1\ldots
a_6 a_7a_8}
\eqno(2.14)
$$
to become 
$$
[R_2^{a_1\ldots a_6}, R_2^{a_7a_8}]={1\over 2}R_1^{a_1\ldots
a_6 a_7a_8}+\tilde R_1^{a_1\ldots a_6 [a_7,a_8]}.
\eqno(2.15)
$$
The new generator $\tilde R_1^{a_1\ldots a_6 a_7,a_8}$ is anti-symmetric
in its first seven indices. The Jacobi identities then imply that this
generator then also appears on the right-hand side of the relations 
$$ 
[R_1^{a_1\ldots a_6}, R_1^{a_7a_8}]=-{1\over 2}R_1^{a_1\ldots
a_6 a_7a_8}+\tilde R_1^{a_1\ldots a_6 [a_7,a_8]}.
\eqno(2.16)
$$
and 
$$
[R_2^{a_1\ldots a_4}, R_2^{a_5\ldots a_8}]= 8\tilde
R_1^{a_1 a_2 a_3 a_4[a_5 a_6 a_7,a_8]}.
\eqno(2.17)
$$
Re-evaluating 
the commutators of equations (2.12) and (2.13) with these modified
relations we now find  the missing  generators of $E_8$ which are
given by 
$$ 
S_{kl}= {1\over 4!}\epsilon _{kl i_1\ldots i_4} R_2^{i_1\ldots i_4},\ 
S_{k 10}= {2\over 5!}\epsilon _{k i_1\ldots i_5} R_2^{i_1\ldots i_5 10},\
$$
$$S_{k 11}= -{2\over 5!}\epsilon _{k i_1\ldots i_5} R_1^{i_1\ldots i_5
10},\ S_{10 11}= -{1\over 6!}\epsilon _{i_1\ldots i_6}
\tilde  R_1^{i_1\ldots i_6 10,10}, 
\eqno(2.18)$$
and 
$$S^k= -{4\over 5.5!}\epsilon _{i_1\ldots i_6}\tilde R^{i_1\ldots i_510
[k, i_6]},\ 
S^{10}= -{4\over 6!}\epsilon _{i_1\ldots i_6} R_1^{i_1\ldots i_6},\ 
S^{11}= -{4\over 6!}\epsilon _{i_1\ldots i_6} R_2^{i_1\ldots i_6}
\eqno(2.19)$$
\par
For the restriction of the restriction of the $G_{IIB}$
algebra considered above that there are no other generators except those
considered so far and these generate the Borel subalgebra of $E_8$. 
The 248 adjoint of $E_8$ decomposes into SL(8,R)
representations as 
$$248 = 1(\sum _{\hat i}K^{\hat i}{}_{\hat i})+63 (K^{\hat i}{}_{\hat j})+
56(X^{\hat i_1\hat i_2\hat i_3})
+\bar {28}(S_{\hat k\hat l}) +\bar {8}(S^{\hat k}) 
\eqno(2.20)$$
as well as the negative roots $\bar{56}+ 28 + 8$. 
\par
As explained in reference [11], the introduction of  the extra
generator $\tilde R_1^{a_1\ldots a_6 a_7,a_8}$ in the $G_{IIB}$
algebra  implies the presence of an additional field $h_{a_1\ldots
a_7,a_8}$  which together with
$h_a{}^b$ provides a first order formulation of gravity.

%%%%%%%%%%%%%%%%%%%%%%%%%%%%%%%%%%%%%%%%%%%%%%%%%%%%%%%%%%
\medskip 
{\bf 3. Discussion}
\medskip
%%%%%%%%%%%%%%%%%%%%%%%%%%%%%%%%%%%%%%%%%%%%%%%%%%%%%%%%%%

In this paper we have shown that the bosonic sector of the IIB
supergravity theory can be formulated as a non-linear realisation 
 of an infinite dimensional algebra
which is the closure of the conformal algebra  and the algebra denoted
above by $G_{IIB}$. This formulation includes the Borel subalgebra of
SL(11) and $E_7$, but we argue that the non-linear realisation can be
enlarged to include the Kac-Moody algebra $E_{11}$. We carry out the
first step in this enlargement and show, by using a first order
formulation of gravity involving two fields which are related by duality,
that  the algebra  contains the Borel subalgebra of
$E_8$. 
\par
It was perhaps not too surprising that the IIA supergravity theory 
has the same Kac-Moody algebra underlying it as the eleven dimensional 
supergravity theory as they are related by a reduction on a circle.
However, IIB supergravity can not be obtained from eleven dimensional 
supergravity in a simple way and so the appearance of the same algebra
is perhaps surprising.  It is consistent with the idea expressed in
[10,11]  that M theory has an underlying $E_{11}$ symmetry and that the
maximal supergravity theories in eleven and ten dimensions appear as 
different manifestations of this symmetry.  
It is instructive to recall that non-linear realisations typically arise
when a symmetry is spontaneously broken and it describes  the theory
controlling the low energy excitations. The local subgroup in the
non-linear realisation corresponds to that part of the original symmetry
that is preserved in the symmetry breaking.  If a theory has different
possible vacua  one finds corresponding different low energy theories
based on the same symmetry group, but with different local symmetry
groups. This is precisely the picture we find with the maximal
supergravities, they possess the  same underlying
group $E_{11}$, but have different local subgroups and so can be
interpreted as different vacua of one theory, namely  M theory. Indeed,
from this perspective what distinguishes the IIA and IIB theory is the
way the SL(11) subgroup is embedded in
$E_{11}$ and correspondingly what parts of it are in the local subgroups.
The embedding is fixed by the occurance of the momentum generator which in
turn gives rise to the space-time coordinates in the theory. 
\par 
The different SL(11) 
embeddings have an  SL(9) subgroups in common. This is  
consistent with the fact that IIA and IIB supergravity theories are the
same  when reduced to nine dimensions where the different embeddings give
rise to the T duality transformations [12] between these two theories.

%%%%%%%%%%%%%%%%%%%%%%%%%%%%%%%%%%%%%%%%%%%%%%%%%%%%%%%%%%%%
\medskip
{\bf Acknowledgements:}
\medskip
%%%%%%%%%%%%%%%%%%%%%%%%%%%%%%%%%%%%%%%%%%%%%%%%%%%%%%%%%%%%

IS would like to thank Andr\'e Miemiec, who has given support when 
calculating the equations of motion for various SUGRAs. IS is financially
supported by DAAD (D/00/09914). PW would like to thank Matthias Gaberdiel 
for discussions.

%%%%%%%%%%%%%%%%%%%%%%%%%%%%%%%%%%%%%%%%%%%%%%%%%%%%%%%%%%%%%
\medskip
{\bf {References}}
\medskip
%%%%%%%%%%%%%%%%%%%%%%%%%%%%%%%%%%%%%%%%%%%%%%%%%%%%%%%%%%%%%

\parskip 0pt

\item{[1]} W. ~Nahm, {\it "Supersymmetries and their representations"},
     Nucl. Phys. {\bf B135} (1978), p.149
\item{[2]} E.~Cremmer, B.~Julia, and J.~Scherk, {\it ``Supergravity
theory in 11 dimensions''},  Phys. Lett. {\bf B76} (1978) 409--412.
\item{[3]}
I.~C.~G. Campbell and P.~C. West, {\it ``N=2 d = 10 nonchiral supergravity and
its spontaneous compactification''},  Nucl. Phys. {\bf B243} (1984)
112.;
  M. Huq and M. Namazie,
{\it ``Kaluza--Klein supergravity in ten dimensions''},
Class.\ Q.\ Grav.\ {\bf 2} (1985).;
  F. Giani and M. Pernici,
{\it ``$N=2$ supergravity in ten dimensions''},
Phys.\ Rev.\ {\bf D30} (1984) 325. 
\item{[4]} J.~H. Schwarz and P.~C. West, ``Symmetries and transformations
of chiral {N}=2,{D} = 10 supergravity,''  Phys. Lett. {\bf B126}
(1983) 301.
\item{[5]} P.~S. Howe and P.~C. West, ``The complete {N}=2, d = 10
supergravity,''  Nucl. Phys. {\bf B238} (1984) 181.
\item{[6]} J.~H. Schwarz, ``Covariant field equations of chiral {N}=2 {D}
= 10 supergravity,'' Nucl. Phys. {\bf B226} (1983) 269.
\item{[7]} E. Cremmer and B. Julia,
{\it ``The $N=8$ supergravity theory. I. The Lagrangian''},
Phys.\ Lett.\ {\bf 80B} (1978) 48
\item{[8]} C.M. Hull and P.K. Townsend,
{\it ``Unity of superstring  dualities''},
Nucl.\ Phys.\ {\bf B438} (1995) 109, hep-th/9410167.
\item {[9]} E.~Cremmer, B.~Julia, H.~L{\"u}, and C.~N. Pope,
  {\it ``Dualisation of dualities. {I}{I}: Twisted self-duality of
    doubled fields and superdualities''},  Nucl. Phys. {\bf B535}
  (1998) 242, {\tt hep-th/9806106}
\item{[10]} P.~C. West, {\it ``Hidden superconformal symmetry in {M}
    theory ''},  JHEP {\bf 08} (2000) 007, {\tt hep-th/05270}
\item{[11]} P. ~West, {\it ``E(11) and {M} theory,''} {\tt hep-th/0104081}
\item{[12]} E.~Bergshoeff, C.~Hull, and T.~Ortin, {\it ``Duality in
    the type {I}{I} superstring effective action''},  Nucl. Phys.
  {\bf B451} (1995) 547--578, {\tt hep-th/9504081}
\item{[13]} P.~West {\it "Supergravity, Brane Dynamics and String
    Duality"}, {\tt hep-th/9811101}

\end